\documentclass[aps,prl,preprint,superscriptaddress,showpacs]{revtex4-1}

\usepackage{graphics}
\usepackage{graphicx}
\usepackage{amsfonts}
\usepackage{textcomp}
\usepackage{amssymb}
\usepackage{amsmath}
\usepackage{multirow}

\begin{document}

\title{Density of States for a Specified Correlation Function \\ and the Energy Landscape}

\author{C. J. Gommes} \email{cedric.gommes@ulg.ac.be}
\affiliation{Department of Chemical Engineering, University of Li\`ege, Li\`ege 4000, Belgium}
\affiliation{Department of Chemistry, Princeton University, Princeton New Jersey 08544, USA}

\author{Y. Jiao} \email{yjiao@princeton.edu}
\affiliation{Department of Chemistry, Princeton University, Princeton New Jersey 08544, USA}

\author{S. Torquato} \email{torquato@electron.princeton.edu}
\affiliation{Department of Chemistry, Princeton University, Princeton New Jersey 08544, USA}
\affiliation{Program in Applied and Computational Mathematics and Princeton Center for Theoretical Science,
Princeton University, Princeton New Jersey 08544, USA}

\date{\today}

\begin{abstract}
The degeneracy of two-phase disordered microstructures consistent with a specified correlation function is analyzed by mapping it to a ground-state degeneracy. We determine for the first time the associated density of states via a Monte Carlo algorithm. Our results are described in terms of the roughness of the energy landscape, defined on a hypercubic configuration space. The use of a Hamming distance in this space enables us to define a roughness metric, which is calculated from the correlation function alone and related quantitatively to the structural degeneracy. This relation is validated for a wide variety of disordered systems.
\end{abstract}

\pacs{05.20.-y, 61.43.-j}

\maketitle

Spatial correlation functions are fundamental descriptors that arise in a variety of disciplines, including condensed matter physics \cite{Zallen:1983}, geostatistics \cite{Chiles:1999}, computer vision and image analysis \cite{Serra:1982}, statistical physics \cite{Chandler:1987}, and materials science \cite{Torquato:2000,Sahimi:2003}. They notably provide a very general tool for characterizing the microstructure of materials and relating this information to their physical properties \cite{Torquato:2000}. Moreover, most experimental techniques available for {\it in situ} studies with a nanometer resolution -- notably scattering methods -- yield information in the form of two-point correlation functions \cite{Feigin:1987,Drake:1991,Barral:1992,Filipponi:1995}.

It is well known that two-point statistics are generally not sufficient to characterize fully a microstructure \cite{Matheron:1975}. This is referred to as the phase problem in crystallography. The ambiguity of two-point information has been investigated theoretically from the perspective of crystallography \cite{Patterson:1944,Hosemann:1954}, computer vision \cite{Aubert:2000}, materials science \cite{Jiao:2009},
and cosmology \cite{Jiao:2009}. In some very special cases, distinct microstructures with identical correlation functions can be derived analytically \cite{Jiao:2010A,Jiao:2010B}. It has also been shown that the structural ambiguity is considerably larger for a radial function without angular information, which is the only data available from small-angle scattering experiments. However, if angular information is employed successful microstructure reconstructions can often be obtained \cite{Hauptman:1986, Rozman:2002, Hansen:2005, Fullwood:2008}.

In the present paper, we determine for the first time a general means to numerically calculate the number of microstructures consistent with {\it any} specified correlation function for arbitrary systems. For concreteness, the present analysis is focused on two-phase microstructures which are suitable models for a host of natural and synthetic materials such as composites \cite{Torquato:2000,Sahimi:2003}, colloidal suspensions and microemulsions \cite{DeGennes:1992}, porous materials \cite{Barton:1999}, etc. Moreover, the general method is applied to the radial two-point correlation function $S_2(r)$ defined as the probability that two random points at distance $r$ from one another both belong to a given phase.

\begin{figure} \label{fig:Example_Reconstruction}
\begin{center}
\includegraphics[width=8.6cm, keepaspectratio]{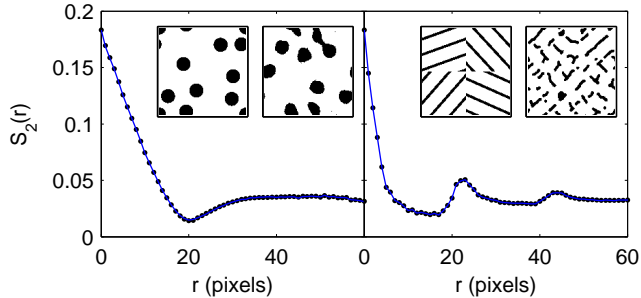}
\end{center}
\caption{Degenerate microstructures corresponding to a hard-disk (left) and a polycrystal (right) two-point correlation functions : in each case, the correlation functions of the two degenerate microstructures ($\bullet$ and $-$) are indistinguishable on the scale of the figure.}
\end{figure}

The general methodology consists in mapping the determination of the microstructure ambiguity to the determination of a ground-state degeneracy. This is achieved in the general framework of reconstruction methods, which aim at producing microstructures consistent with a target correlation function $\hat S_2(r)$. In that context, a microstructure having a correlation function $S_2(r)$ is associated with an ``energy'' defined as \cite{Yeong:1998}
\begin{equation} \label{eq:definition_E}
E = \sum_r \left[ \hat S_2(r)-S_2(r) \right]^2,
\end{equation}
which is equivalent to a norm-2 error. All the microstructures consistent with a specified $\hat S_2(r)$ are the ground states
(with globally minimized energy) of the corresponding reconstruction problem. Examples of degenerate microstructures, obtained via simulated annealing \cite{Yeong:1998,Jiao:2008}, are given in Fig.~1 for correlation functions typical of hard-disk and polycrystal microstructures. In both cases the two displayed microstructures have the same two-point correlation function to within an error of $10^{-7}$, but their higher-order correlation functions differ.

Using the language of solid-state physics, we refer to the number $\Omega(E)$ of microstructures having energy $E$ as the density of states (DOS). An efficient method for estimating the DOS has been proposed by Wang and Landau
\cite{Wang:2001A,Wang:2001B}, and applied to a wide variety of problems ranging from solid-state physics \cite{Yamaguchi:2001}, to biophysics \cite{Rathore:2002}, and logic \cite{Ermon_2010}. The algorithm is based on the observation that a Monte Carlo (MC) move from state $i$ to $j$ with acceptance probability
\begin{equation} \label{eq:Wang_Landau}
p_{i \to j} = \min \left\{1, \ \Omega(E_i) / \Omega(E_j) \right\}
\end{equation}
would lead the system to visit all energies with the same probability. Because $\Omega(E)$ is unknown, the DOS is initialized to $\Omega(E) = 1$ for all energies, and the MC algorithm updates this value until convergence is achieved.

We restrict the discussion to discrete two-phase microstructures, which can be thought of as an image composed of black and white pixels. Starting from an initial configuration with energy $E_i$, a black pixel is moved randomly to an unoccupied pixel. The correlation function is updated, the new energy $E_f$ is calculated through Eq. (\ref{eq:definition_E}), and the move is accepted or rejected according to Eq. (\ref{eq:Wang_Landau}). Each time a given energy is visited, a histogram is updated, $H(E) \to H(E) +1$, and the estimated density of states is updated according to $\Omega(E) \to F \times \Omega(E)$ where $F$ is a numerical factor larger than 1. The evolution continues with the updated value of $\Omega(E)$ until the histogram $H(E)$ is flat. At this point $H(E)$ is reset to $0$, $F$ is reduced to $\sqrt{F}$, and the evolution starts over again. The entire procedure is repeated until $F$ becomes lower than a prescribed accuracy.

\begin{figure}
\begin{center}
\includegraphics[width=8.6cm, keepaspectratio]{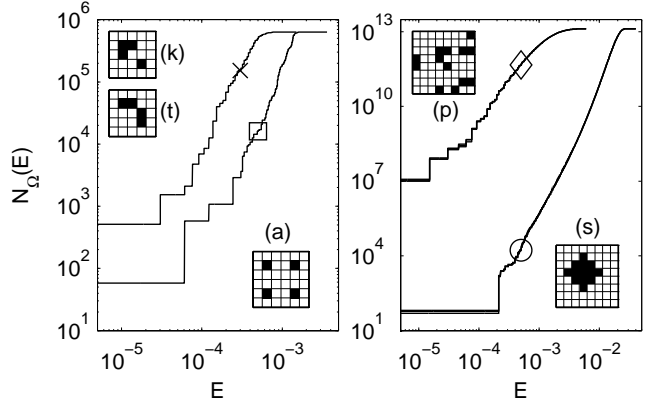}
\end{center}
\caption{Cumulative densities of states for various correlation functions: morphologies (a) and (s) are uniquely determined by their correlation functions (${\square}$ and $\circ$); morphologies (k) and (t) have identical correlation functions ($\times$); morphology (p) is a realization of a Poisson process ($\diamond$). Each curve is the superposition of 3 independent Monte Carlo runs.}
\label{fig:DOS}
\end{figure}

Examples of DOS calculations are given in Fig. \ref{fig:DOS}. The results are given in terms of the cumulative DOS
\[
N_\Omega(E) = \sum_{e\leq E} \Omega(e)
\]
which has been normalized for $N_\Omega(E \to \infty)$ to be equal to the total number of configurations $\Omega_{tot} = \binom{N}{N_1}$, with $N$ the total number of pixels and $N_1$ the number of black pixels. The limit $N_\Omega(E \to 0)$ is the ground-state degeneracy $\Omega_0$, i.e. the total number of microstructures consistent with $\hat S_2(r)$.

The ``square'' (a) and ``sphere'' (s) microstructures in Fig. \ref{fig:DOS} are uniquely specified by their correlation functions. The degeneracy $\Omega_0$ is therefore only a trivial contribution resulting from the $64$ possible translations on a $8 \times 8$ grid. The values found from the MC estimations are $66 \pm 7$ and $ 58 \pm 8$, for (a) and (s) respectively, with the error estimated from 3 independent runs. Configurations (k) and (t) are the ``Kite \& Trapezoid'' configurations discussed in a previous paper \cite{Jiao:2010A}; they have identical correlation functions, which results in a non-trivial factor 2 to the ground-state degeneracy. Moreover, as the configurations are lacking rotational symmetry, the possible orientations contribute a trivial factor 4. The ground state degeneracy is therefore $\Omega_0 = 512 $, in excellent agreement with the MC estimate $500 \pm 68$. Finally, configuration (p) is a realization of a Poisson point process \cite{Serra:1982} used here as a model of a very disordered morphology. This particular microstructure is found to be extremely degenerate with $\Omega_0$ as large as  $ (11 \pm 1) \ 10^6 $.

\begin{figure}
\begin{center}
\includegraphics[width=4.5cm, keepaspectratio]{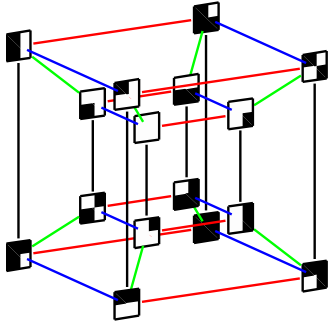} \\
\end{center}
\caption{The configuration space $\mathcal{C}$ of a two-phase morphology is a $N$-dimensional hypercube, on which a Hamming distance is defined. Each dimension is associated with the phase of a particular pixel. For a $2 \times 2$ morphology, $\mathcal{C}$ is a tesseract, with each 4-dimensional direction represented here by a color.}
\label{fig:Hypercube}
\end{figure}

It is generally acknowledged that large ground-state degeneracy is a characteristic of systems having a rough energy landscape \cite{Wales:1998}. We observe that the configuration space of discrete two-phase microstructures is the set of vertices of a $N$-dimensional hypercube (see Fig. \ref{fig:Hypercube}). This results from the properties of the indicator vector $I(i)$, with components equal to $1$ when point $i$ is a black pixel and $0$ otherwise, which can be seen as a coordinate in $N$-dimensional space. Moving the system along a given $N$-dimensional direction corresponds to swapping a particular pixel between black and white. Once a target correlation function $\hat S_2(r)$ is specified, each vertex is associated with an energy $E$ according to Eq. (\ref{eq:definition_E}).

The roughness of the energy landscape characterizes the spatial variability of $E$ in configuration space. It is meaningful therefore to define a distance in this configuration space. A natural choice is the {\it Hamming} distance, which counts the number of edges between any two vertices. Interestingly, if the number of black pixels is kept constant, i.e. $ \sum I(i) = N_1 $, all the realizable microstructures lie on the intersection of the hypercube with a hyperplane. Since the Hamming distance within the hyperplane takes only even values, the distance $d[A,B]$ between two microstructures $A$ and $B$ is defined as half the Hamming distance
\[
d[A,B]=\frac{1}{2} \sum_{i=1}^N \left(I_A(i)-I_B(i) \right)^2
\]
In real space, the distance $d$ is the smallest number of black-pixel displacements required to pass from $A$ to $B$.

\begin{figure}
\begin{center}
\begin{tabular}{c}
\includegraphics[width=8.6cm, keepaspectratio]{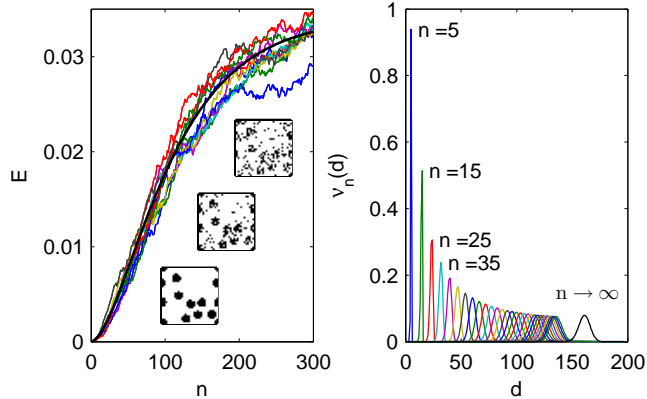}
\end{tabular}
\end{center}
\caption{Left: Energies visited during a random walk in configuration space, starting from a particular ground state of a hard-disk correlation function. Each color is a realization and the black line is the average calculated analytically. Examples of microstructures visited for $n = 50 $ and $n = 100$ are displayed. Right : Statistical distribution $\nu_n(d)$ of the distances $d$ from the ground state of the morphologies visited after an increasing number of steps $n$ of the random walk.}
\label{fig:RA_example}
\end{figure}

To address analytically the question of the roughness of the energy landscape, we use a {\it random walk in configuration space}, as illustrated in Fig. \ref{fig:RA_example}. Starting from a ground state, i.e. a microstructure with $E=0$, the system makes successive random jumps of length $d=1$ in configuration space $\mathcal{C}$. When the number of jumps $n$ increases, the random walk explores the configuration space over increasingly large distances $d$ from the starting microstructure. The rate at which the energy increases with $n$ characterizes the energy landscape.

An analytical expression is derived in the Supplemental Material \cite{SupportingInformation}
for the average energy, and for the statistical distribution of distances to the ground state, both as a function of $n$. Combining these two pieces of information yields a characteristic energy profile for the basin of any ground state. A detailed analysis is provided elsewhere \cite{Gommes:2011}, and we focus here on two particular values on the average energy curve. The first is the average energy $\left< E \right>$ in the limit $n \to \infty$ and the second is the average energy $E(1)$ reached for $n =1$. The random walk being ergodic, $\left< E \right>$ is equal to the average energy of all $\Omega_{tot}$ possible states. The other value, $E(1)$, is a measure of the curvature of the energy landscape near the starting ground state. This results from the observation that $E(1)$ is the average energy of all states at distance $d =1$ from the ground state, the latter having zeros energy. The ratio $E(1)/\left< E\right>$ therefore provides a metric for the local roughness of the energy landscape.

The full mathematical expressions for $E(1)$ and $\left< E \right>$ can be found in the Supplemental Material \cite{SupportingInformation}. The quantity $\left< E \right>$ is a global characteristic of the energy landscape, which accordingly depends only on $\hat S_2(r)$. By contrast, $E(1)$ depends also on the particular ground state used as the starting point for the random walk via a particular function $\sigma^2_C(r)$ with the following structural meaning. Imagine choosing randomly a black pixel in the ground state and drawing a circle of radius $r$ centered on it (or a sphere in $3D$). The fraction of the circle that overlaps a black pixel is a random variable $\varphi_r$ that depends on the particular black pixel chosen as the center. The variance of $\varphi_r$ is the function $\sigma^2_C(r)$, which can be thought of as a generalized ``coarseness'' \cite{Lu:1990}.

The function $\sigma^2_C(r)$ carries 3-point structural information in excess to $\hat S_2(r)$. Thus, $\sigma^2_C(r)$ could differ significantly from one ground state to another. In practice, however, the asymptotic behavior of $\sigma^2_C(r)$ for both large and small $r$ can be expressed in terms of $\hat S_2(r)$ alone, which enables us to derive a single approximation $\tilde \sigma^2_C(r)$ common to all ground states \cite{SupportingInformation}. Using that approximation yields a single value for $E(1)/\left< E  \right>$, which is a global metric of the roughness of the energy landscape.

\begin{figure}
\begin{center}
\includegraphics[width=7.5cm, keepaspectratio]{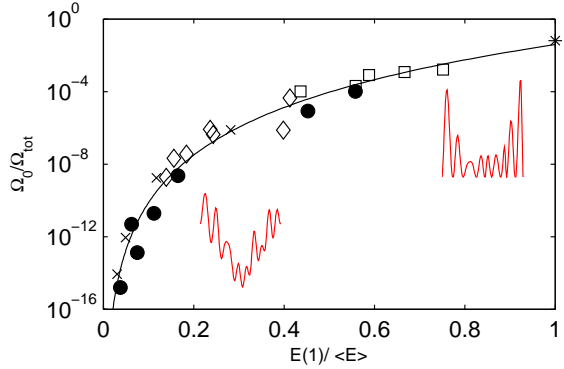}
\end{center}
\caption{Relation between ground-state degeneracy $\Omega_0/\Omega_{tot}$ and roughness of the energy landscape $E(1)/\left< E \right>$. The various microstructures are: disks of different sizes ($\bullet$), realizations of Poisson processes ($\diamond$), hard-disk systems ($\times$), 4-point configurations ($\square$), and 2-point configurations ($*$). The black line is a guide to the eye; the insets are sketches of the energy landscape for large and small values of $E(1)/\left< E \right>$.}
\label{fig:Roughness}
\end{figure}

Figure \ref{fig:Roughness} shows the quantitative relation between $E(1)/\left< E\right>$ and a normalized ground state degeneracy $\Omega_0/\Omega_{tot}$, for a variety of microstructures defined on a $8 \times 8$ grid, with values of $N_1$ ranging from 2 to $32$ (see Supplemental Material \cite{SupportingInformation}). They comprise a series of objects uniquely characterized by their correlation functions, which have therefore only a trivial degeneracy, as well as a series of non-trivially degenerate realizations of Poisson point processes. In each case, the ground-state degeneracy was estimated via the MC algorithm, and the roughness metric was calculated from $\hat S_2(r)$ using the approximation $\tilde \sigma^2_C(r)$. The ratio $\Omega_0/\Omega_{tot}$ is found to be {\em highly} correlated with the ratio $E(1)/E_\infty$ over more than 15 orders of magnitude.

When passing from small to large values of $E(1)/\left< E \right>$, the energy landscape changes qualitatively in the way suggested by the insets in Fig. \ref{fig:Roughness}. For low values of $E(1)/\left< E \right>$, the energy landscapes has an overall funnel structure, with low-energy barriers, which makes it well suited for optimization problems. By contrast, for large values of $E(1)/\left< E \right>$, the landscape is very rough with a large number of ground states. However, it is interesting to note that the rightmost point in Fig. \ref{fig:Roughness} is obtained for a system with $N_1 = 2$ having thus only a trivial degeneracy. The corresponding energy landscape is extremely rough because any possible energy can be found at a distance as short as $d = 1$ from the ground state, but the total number of configurations $\Omega_{tot}$ is also extremely small.

The data referred to as disks in Fig. \ref{fig:Roughness} is a collection of non-degenerate microstructures with increasing values of $N_1$. As $N_1$ increases, the roughness $E(1)/\left< E \right>$ decreases but the degeneracy remains equal to its trivial translation contribution $\Omega_0 = N$. Interestingly, the values of $\Omega_0/\Omega_{tot}$ of these non-degenerate microstructures span the same curve as the realizations of Poisson processes, for which $\Omega_0$ has a huge non-trivial contribution. The observation that the roughness-degeneracy relation does not discriminate trivial from non-trivial degeneracies suggests ways to determine analytically the degeneracy of large microstructures, which are out of the reach of MC methods \cite{Gommes:2011}.

To summarize, the long-standing problem of the degeneracy of microstructures compatible with a specified two-point correlation function can be tackled obliquely through the characterization of its associated energy landscape. The results of our MC calculations show that the roughness of the energy landscape is indeed highly correlated with the ground state degeneracy. The MC algorithm converges only for very small systems \cite{Dayal:2004} but the roughness metric we derived, $E(1)/ \left< E \right>$ , can be calculated analytically from $\hat S_2(r)$ with no limit of size or dimensionality.

Our results have ramifications in the manifold of fields where correlation functions are useful. In the particular context of materials science, they may contribute to topics as diverse as the understanding of the sharpness of variational bounds for materials properties \cite{Torquato:2000} and the quantitative analysis of scattering experiments \cite{Feigin:1987}. In computer vision, they may help assess the robustness of second-order texture classifiers \cite{Serra:1982}. They also have direct applications throughout many subfields of physics because $\Omega_0$ is a lower bound for the physical ground-state degeneracy of any system with pairwise potential energy \cite{Gommes:2011}.

Our general methodology can be applied to correlation functions other than $S_2(r)$; this includes higher-order \cite{To83} as well as cluster correlation functions \cite{Jiao:2009,Zachary:2011}. More generally,  random walks in configuration space can be used to derive roughness metrics for the energy landscape of any physical problem, notably protein folding \cite{Chavez:2004}, complex chemical reactions \cite{Wales:1998}, phase equilibria in nanopores \cite{Puibasset:2010}, and glass transition \cite{Debenedetti:2001}. For instance, the modeling of spin glasses with frustrated Ising models \cite{Parisi:2006} yields an energy that is a linear functional of $S_2(r)$, which is comparatively simpler than the quadratic functional we considered here. We hope to investigate this in future work.

\begin{acknowledgments}
C.J.G. acknowledges support from the Fonds de la Recherche Scientifique (F.R.S.-FNRS, Belgium); S.T. and Y.J. were supported by the Office of Basic Energy Science, Division of Materials Science and Engineering under Award DE-FG02-04-ER46108.
\end{acknowledgments}


%

\end{document}